# Physics and Applications of Laser Diode Chaos
M. Sciamanna[1,*] and K. A. Shore[2]

*Abstract – An overview of chaos in laser diodes is provided which surveys experimental achievements in the area and explains the theory behind the phenomenon. The fundamental physics underpinning this behaviour and also the opportunities for harnessing laser diode chaos for potential applications are discussed. The availability and ease of operation of laser diodes, in a wide range of configurations, make them a convenient test-bed for exploring basic aspects of nonlinear and chaotic dynamics. It also makes them attractive for practical tasks, such as chaos-based secure communications and random number generation. Avenues for future research and development of chaotic laser diodes are also identified.*

The emergence of irregular pulsations and dynamical instabilities from a laser were first noted in the very early stages of the laser development. Pulses whose amplitude "vary in an erratic manner" were reported in the output of the ruby solid-state laser[1] [Fig. 1 a] and then found in numerical simulations[2]. However the lack of knowledge of what would later be termed "chaos" resulted in these initial observations being either left unexplained or wrongly attributed to noise.

The situation changed in the late 1960s with the discovery of sensitivity to initial conditions by Lorenz[3], later popularized as the "butterfly effect". As illustrated in Fig. 1 (b), numerical simulations of a deterministic model of only three nonlinear equations showed an irregular pulsing with a remarkable feature: the state variables evolve along very different trajectories despite starting from approximately the same initial values [Fig. 1 b]. The distance $\delta(t)$ between nearby trajectories diverges exponentially : $\delta(t) = \delta(0) \exp(\lambda t)$ provided that $\lambda$, the effective Lyapunov exponent of the dynamical system, is positive. Consequently such systems are unpredictable in the long term. Plotted in the x-y-z phase space of the state variables, the trajectories converge to an "attractor" which has the geometric property of being bounded in space despite the exponential divergence of nearby trajectories [Fig. 1 c]. Such attractors are found to have a fractional dimension[4] and are thus termed strange. Aperiodicity, sensitive dependency to initial conditions and strangeness are commonly considered as the main properties for "chaos"[5]. Numerous practical algorithms are available today to differentiate between deterministic chaos and stochastic noise, including estimating the dominant positive Lyapunov exponent[6] and the fractal dimension[7] as illustrated in Fig. 1 d and Fig. 1 e for Lorenz chaos.

The fields of laser physics and chaos theory developed independently until 1975[8] when Haken discovered a striking analogy between the Lorenz equations that model fluid convection and the Maxwell-Bloch equations modelling light-matter interaction in single-mode lasers. The nonlinear interaction between the wave propagation in the laser cavity (represented by the electric field *E*) and radiative recombination producing macroscopic polarization (encapsulated in the polarization *P* and carrier inversion *N*) yield similar dynamical instabilities to those found in the Lorenz equations. More specifically, in addition to the conventional laser threshold, Haken suggested a second threshold would exist above which "spiking occurs randomly though the equations are completely deterministic"[8].

Motivated by the new theoretical developments of chaos theory[9-11], laser experimentalists started an intense search for Lorenz-Haken chaos. However, the Haken second threshold requires a high loss resonator and pumping the laser at about 10-20 times the first laser threshold. In addition, although the Lorenz and laser Maxwell-Bloch equations share many similar dynamical properties, the parameters that determine the relaxations of the state variables in both physical systems take very different values. In particular in some lasers, the polarization and/or the carrier inversion relax much faster than the field and thus can be adiabatically eliminated from the other equations, hence reducing the system's dimension. The first conclusive experimental reports of laser chaos were therefore obtained in a $CO_2$ laser where loss modulation provides the additional degrees of freedom to achieve chaotic trajectories[12,13]. Lorenz-Haken chaos in a free-running laser was only later achieved in an 81.5 µm-$NH_3$ laser[14], in which low pressure and long wavelength combine to reduce the secondary laser threshold [Fig. 1 f].

**Laser diode: damped nonlinear oscillator**

These early demonstrations motivated investigations in more practical lasers such as laser diodes. Laser diodes have numerous applications including imaging, sensing, fibre-optic communications and spectroscopy. Initially aimed at providing a constant output power, laser diodes are today commonly used to produce periodical short optical pulses at high repetition rates[15]. Besides steady operation and pulsing dynamics, chaos theory reveals that a nonlinear physical system of high enough dimension may bifurcate to more complex dynamics including chaos. This applies also to laser diodes. However in considering the analogy with Lorenz chaos, it must be appreciated that in laser diodes the polarization typically relaxes much faster (at $\gamma_P$ rate) than the field (at $\kappa$ rate) and the carrier inversion (at $\gamma_N$ rate), i.e. $\gamma_P \gg \kappa > \gamma_N$ [16]. As a result, laser diode dynamics is described using rate equations for the field and carrier inversion as a driven damped nonlinear oscillator and is therefore limited to a spiralling flow toward a steady-state (so-called relaxation oscillations) [Fig. 1 g]. However, laser diodes, almost uniquely, possess a property which makes them extremely sensitive to optical perturbations: their emission frequency is detuned from the gain spectrum peak leading to an anomalous dispersion effect at the lasing frequency. This property results in a refractive index variation with carrier density and translates into a so-called $\alpha$ factor that explains laser chirp and linewidth broadening[17] but also facilitates laser instabilities[18,19].

In the following article we review situations in which a laser diode can be brought into chaos and then describe currently identified applications of laser diode chaos. Finally we offer reasons for our expectation that this area of activity will remain a fertile field for future research.

**Configurations for achieving laser diode chaos**

Several configurations may be used to overcome the damped relaxation oscillations and therefore to generate laser diode chaos.

*External optical feedback*
Returning a small fraction of the laser emission into the laser diode cavity may result in a chaotic output with different types of waveforms and properties[20] [Fig. 2 a]. The richness of the dynamics results from the competition between the laser's intrinsic relaxation oscillation frequency $f_{RO}$ and the frequency $f_{EC} = c/(2L)$ of the external cavity, where $L$ is the distance from the laser to the external optical feedback element[21,22]. Chaotic dynamics in the case $f_{EC} \gg f_{RO}$ include low-frequency power fluctuations[23] or coherence collapse dynamics[24] typically giving a high dimensional chaotic attractor[25]. By contrast, for $f_{EC} \ll f_{RO}$ different self-organizing dynamics take place including periodic pulsing at the external-cavity frequency[22], with bifurcations to quasiperiodic dynamics such as regular pulse packages[26].

The feedback can be provided either from a simple external mirror or with more complex configurations leading to different routes to chaos, e.g. feedback from a phase-conjugate mirror[27,28] or including a Fabry-Perot etalon as a filter in the external cavity[29]. Numerous works have extended the study to multimode laser dynamics by considering e.g. the inclusion of diffraction grating in the external cavity[30], or polarizers[31] or retarding plates[32] as ways to influence the polarization of the returning field; thus providing additional ways to select and control specific pulsing or chaotic dynamics[33].

External optical feedback is the most prominent configuration in applications using laser diode chaos and commonly used for chaos secure communications, random number generation, chaos computing and sensing. The scaling of the time-delay with respect to the laser internal time-scale and the sensitivity to the phase of the returning field produce means to induce various dynamical scenarios leading to chaos[34,35] and to engineer high-dimensional chaotic waveforms spanning over large frequency bandwidth.

*Optical injection (unidirectional coupling)*
Optical injection from another laser can also be used to de-stabilise lasers [Fig. 2 b]. In situations of large frequency detuning and/or strong injection, the injected laser destabilizes to chaos through different bifurcation mechanisms[36,37]. The availability of mathematical continuation techniques that follow bifurcations in a two-dimensional parameter plane together with the limited set of parameters influencing the nonlinear dynamics has made it possible to reach an unprecedented global agreement between experiment and theory over a large range of injection parameters[37]. By harnessing the optical injection nonlinear dynamics, one can engineer and select dynamics for specific needs. The transition to chaos through period doubling has been used for example as a frequency multiplication or conversion processes in photonic microwave generation[38] or in remote sensing where an even weak irradiance injecting light into the diode laser is detected through the resulting bifurcations[39]. As with optical feedback, injection of either a polarized field or a single mode into a laser emitting in several polarizations[40] or longitudinal mode[41] components leads to new mechanisms for chaos instabilities but simultaneously provides additional ways to control the dynamics.

*External current modulation*
Direct current modulation of a laser diode [Fig. 2 c] is a common practice. Less familiar is the opportunity to achieve chaotic pulsing when the modulation frequency is close to the laser relaxation oscillation frequency and/or the modulation depth is relatively large. The first theoretical works suggested a period doubling route to chaos with increased modulation depth[42]. Experiments showed however that the intrinsic noise from quantum fluctuations typically prevents the observation of a period doubling cascade to chaos[43], unless there is careful tuning of both modulation frequency and depth[44]. Recent works have indicated that specific laser diodes such as VCSELs whose dynamics may involve several polarization or transverse modes[45] show chaos for a much wider range of the modulation parameters.

*Loss modulation using saturable absorber*

Modulation of the optical losses in the laser resonator can also cause dynamical instabilities. This is typically achieved by combining a gain section with a reverse bias section that behaves as a saturable absorber [Fig. 2 e]. A noisy spike whose intensity is large enough to saturate the absorber, will initiate a process of loss modulation with the periodicity of the round-trip propagation of the pulse – so called passive mode-locking. Increasing either the reverse bias voltage in the absorber section or the current in the gain section destabilizes the self-pulsing leading to period-doubling (harmonic mode locking) and further to chaos[46]. Although harmonic passive mode locking has been observed experimentally[47], experimental observation of chaotic pulsing has been limited to two-section quantum dot laser diodes where the carrier dynamics has additional features[48].

Under positive bias in both sections, another mechanism for the instability of self-pulsations arises. The laser diode $\alpha$ factor couples any change of the differential gain (as induced by variation of the current) to a change of the refractive index thus reducing the wavelength detuning between the two sections. When the current in one of the two sections is large enough this detuning cannot be compensated. The system then behaves like two coupled but strongly detuned nonlinear oscillators showing chaos[49,50].

Narrow-stripe semiconductor lasers used in optical storage (CD lasers) also exhibit self-pulsing instabilities [Fig. 2 f]. Two blocking layers confine the current injection to the centre of the device, leaving unpumped regions at either side. The penetration of the optical mode into these unpumped regions creates saturable absorption that transversally modulates the optical losses. The so-called Yamada model[51] has clarified the onset of self-pulsation. Chaotic self-pulsations in these devices have been experimentally observed with the addition of external modulation[52] or external optical feedback[53].

*Opto-electronic feedback*

In opto-electronic feedback the output of the laser diode is first converted to electrical current by a photodiode before being amplified and re-applied via the laser driving current [Fig. 2 d]. In this way the laser diode experiences a time-delayed contribution to its dynamics. The feedback signal is termed "incoherent" since it only interacts with the carriers. The feedback may be positive or negative depending on the polarity of the amplifier in the feedback loop. Nonlinear dynamics typically arise when the time-delay is larger than the laser relaxation oscillation time-period. Experiments have captured regular pulsing dynamics but also quasiperiodic bifurcations to chaos when varying the time-delay and/or the feedback strength[54]. The parameter region showing chaos is wider in the case of the negative opto-electronic feedback[55].

*Integrated on-chip chaotic laser diode*

Several proposals have been made for integrating optical feedback leading to on-chip chaotic laser diodes. Three-contact laser diodes have been developed integrating a DFB lasing section with a gain or absorber section and a passive waveguide providing integrated optical feedback with short external cavity (200 μm)[56] [Fig. 2 g]. The current driving the gain/absorber section enables control of the feedback strength, whilst the current driving the passive waveguide is used for adjusting the feedback phase. Both a quasiperiodic route to chaos[57] and a period doubling route to chaos[58] have been reported. Adding a fourth-contact section for separated phase control in a longer (1 cm) passive waveguide has enabled achievement of a period doubling route to chaos of much higher dimension[59]. In a recent proposal, the linear waveguide is replaced by an architecture based on a ring passive waveguide of about 1cm length that integrates a DFB section, two SOA sections and a photodiode [Fig. 2 i]. The larger feedback strength enables a dynamical regime of strong competition between $f_{RO}$ and $f_{EC}$ such that chaos is achieved with both high dimension and a featureless broadband power spectrum[60].

Combining two DFB lasing sections separated by a passive waveguide [Fig. 2 h], leads to optical chaos in a tandem semiconductor laser[61]. The three contacts allow control of the frequency detuning, injection strength and phase shift of the mutually coupled lasers, hence accessing chaotic dynamics.

*Mode competition*

In some laser diodes emitting in several longitudinal, transverse or polarization modes, a sufficiently strong nonlinear mode coupling may induce chaotic instabilities even in absence of additional parameter modulation, injection or feedback [Fig. 2 j]. For example, VCSELs typically emit in several polarization modes whose selection results from several mechanisms including a nonlinear coupling of circularly polarized field components through carrier spin-flip relaxation mechanisms[62]. This mechanism for VCSEL polarization selection is accompanied by a sequence of bifurcations to polarization chaos of low-dimension[63]. Also broad-area semiconductor lasers may experience strong coupling between spatial transverse modes through carrier diffusion leading to a large variety of complex spatial patterns including wave chaos[64].

*Nonlinear hybrid opto-electronic feedback*

In the previously described configurations, optical chaos is achieved from the nonlinear coupling between the lasing electric field and the carrier density. Another approach, inspired by the work of Ikeda on passive optical cavities with time-delayed feedback[65], involves combining a laser diode with an optoelectronic feedback loop that contains a nonlinear optical device.

The so-called wavelength chaos generator [Fig. 2 k] uses a wavelength-tunable two-electrode DBR laser diode with a birefringent crystal placed between two crossed-polarizers[66]. This nonlinear optical device is used as an interferometer that converts a variation of the wavelength into a variation of light intensity through a nonlinear function. The intensity change is then converted to an electrical current that is delayed and amplified before driving the laser DBR section, hence impacting the wavelength dynamics. The physics underlying chaos is similar to the one resulting from the Ikeda equation for delayed passive Kerr cavities[65].

Another way to modulate an optical interference function is to use a Mach-Zender modulator [Fig. 2 l]. The intensity of the light emitted by a continuous-wave 1.55 µm DFB laser is modulated using the Pockels electro-optic effect in a LiNbO3 crystal in one arm of the interferometer. When the signals recombine at the output, the resulting interference depends on both the constant and fluctuating voltages applied to the two electrodes across the crystal. The feedback loop provides both low-pass and high-pass filtering described by a delayed integro-differential equation[67]. Both experiment and theory demonstrate a variety of new chaotic dynamics not observed in the conventional Ikeda equation. In particular, exploiting the form of the nonlinear interference function leads to chaotic dynamics with exceptionally flat power spectra over a large bandwidth[68].

## Chaos Communications

Several approaches have been adopted to perform digital communications using synchronized chaotic lasers[69]. Due to their direct compatibility with existing optical fibre communications technology, semiconductor lasers have gained widespread attention for use in optical chaos communications.

*Chaos Synchronisation*

The fundamental requirement for performing communications using a chaotic carrier is achieving chaos synchronisation. Work of Pecora and Carroll[70] stimulated the first observation of synchronization in lasers[71] and led to subsequent experimental realisations with, in particular, the first demonstration of chaos synchronisation in external cavity laser diodes[72].

The generic experimental configuration (Figure 3 a) includes a transmitter or master laser and a receiver or slave laser. Uni-directional optical coupling between the transmitter and slave laser enables their synchronisation. The transmitter laser is made chaotic using optical feedback from an external mirror – an external cavity laser. The receiver laser may be configured as an external cavity laser ("closed-loop") or else it may be a stand alone laser whose dynamics is affected via optical coupling from the chaotic transmitter laser ("open-loop").

To clearly demonstrate the achievement of synchronisation a synchronization diagram is obtained (Figure 3 b). For two perfectly synchronized lasers, the synchronization diagram will be a straight line with a positive gradient. Adjustment of both the strength of optical coupling and the frequency detuning between the two laser diodes has been shown to affect the synchronization between the lasers[73,74]. Mapping the synchronization quality in the plane of the coupling parameters unveils two regions of different synchronization properties[75] (Figure 3c). For strong injection, synchronization occurs through nonlinear amplification of the slave laser and the corresponding parameter region is bounded by bifurcations delimiting injection locking in laser diode. Such synchronization is said to be "generalized" since the receiver laser reproduces an amplified version of the transmitter output. At low injection strength, a much narrower synchronization region is found where the receiver emits a replica of the transmitter laser output ("identical" or "complete" synchronization). Varying the detuning, a negative gradient was found in the synchronization diagram and termed inverse synchronization[76] or anti-synchronization[77].

*Lag/lead/contemporaneous synchronisation*

As observed numerically[78] and experimentally[73], in generalized synchronization the slave laser output at a given time synchronises with the master laser output taking account of a lag time arising by the time of flight between the lasers. In complete synchronization, synchronization occurs between the slave laser output and the time-shifted master laser output. The time-shift is determined by the difference between the coupling time and the external-cavity time-delay, as shown numerically[79] and experimentally[80]. Varying the time-delay the slave laser output may therefore even anticipate that of the master laser[80,81]. Control of leader or laggard dynamics has been explored[82] including demonstration of zero-lag long-range synchronisation[83].

*Message Transmission*

The achievement of high-quality chaos synchronisation enables message transmission using a chaotic carrier. The simple concept here is that a message added to a chaotic carrier generated in a transmitter laser can be recovered using a receiver laser in chaotic synchrony to the transmitter laser. Several schemes for message encoding have been explored and are summarized in Figure 3d. They include i)chaos masking –where the message is simply added to the chaotic carrier[78] ; ii) chaos modulation –proposed for Chua's circuits[84] and used in the pioneering work on optical chaos communications by VanWiggeren and Roy[85] iii) chaos shift keying (CSK)[86] where digital '1's and '0's are associated with distinct states and iv) ON-OFF shift keying (OOSK)[87] where the system is synchronized for say a 1 whilst being unsynchronized for a 0. The CSK scheme offers higher security but is more difficult to implement. In addition, the need to be able to identify the defined states adds a latency to the decoding process which will reduce the achievable bit-rate in transmission. A similar limitation arises for OOSK. To achieve high bit-rate chaotic optical communications attention needs to be paid furthermore to the impact of noise which, in particular, may affect the quality of synchronization and even cause de-synchronization[88].

Using such encoding techniques much effort has been directed at laboratory demonstrations of chaos communications using laser diodes. Such demonstrations included effecting network operations such as message relay[89] and message broadcasting[90]. An experimental arrangement for chaos broadcasting is given in Figure 3e. Good chaos synchronisation is achieved between the transmitter and receiver 1 and between the transmitter and receiver 2.

However the most significant experimental achievement in this respect was a field trial which demonstrated that chaos communications could be effected over the 120km metropolitan area network of Athens, Greece[91]. See Figure 3f. This provided a key demonstration of the suitability of the approach for practical deployment over installed fibre-optic communication channels.

### *Multiplexed Chaotic Communications*

A prominent feature of advanced optical communications systems is their capability to multiplex several laser wavelengths – in so-called wavelength division multiplexed (WDM) operation. To effect WDM operation it was shown[92] that the longitudinal modes of two single-mode lasers may be chaos-synchronised to longitudinal modes of a multi-mode laser. Similar multiplexed chaos synchronization was achieved between either transverse modes or polarization modes of VCSELs[93]. Selection of the polarization state of the coupled light in VCSELs allows synchronization of the polarization modes while keeping the total intensities de-synchronized[93].

These approaches to multiplexing chaotic light require however as many lasers or modes as the number of users. A more spectrally efficient approach has been suggested. Multiplexed encryption using chaotic systems with multiple stochastic-delayed feedbacks have been used to experimentally demonstrate data transmission and recovery between multiple users at several Gbits/s on a single communication channel[94]. Alternatively, a specific coupling scheme was suggested where a pair of laser diodes synchronize to their counterpart at the receiver side although the transmitter lasers lase at the same wavelength and their outputs are combined in a single communication channel[95].

### *Optimised Communications and Transmission Security*

To enable effective chaos communications a minimal message strength is required to ensure acceptable Bit Error Rates (BER) in transmission. However use of a very strong message may compromise the privacy of the message transmission. It is possible to identify optimized regimes of operations where those requirements are balanced[96]. The privacy of chaos communications rests on hardware keys and notably the device parameters of the transmitter and receiver lasers which need to be rather closely matched to achieve high-quality synchronization. Although not guaranteed, security is greatly enhanced by either arranging that synchronization is only possible for a narrow range of parameters, or with a chaotic transmitter of high enough complexity to prevent reconstruction from time-series analysis. In this respect, one would favor schemes that make the synchronization extremely sensitive to parameters (e.g. closed loop configuration) or dependent on many degrees of freedom (such as in coupled VCSELs with polarization-dependent injection[93]). Reconstruction of the chaotic attractor from the observed system output requires the identification of the system parameters. For external cavity lasers the time-delay in the external cavity and the optical feedback strengths are key parameters in determining the laser dynamics. It has been shown to be possible to extract the time-delay from the laser dynamics[97] and hence measures are needed to counter-act such a breach of security. Several means for concealing the time-delay value are available : setting the time-delay close to the time-period of the relaxation oscillations[98], using more than one time-delayed feedback[99], using stochastically varying or modulated time-delay values in the feedback loop[94], or exploiting the polarization properties of VCSELs[99]. Optimization of chaos communication parameters also benefits from recent investigations of complexity measures from experimental time-series[100-102]. Besides conventional algorithms that compute Lyapunov exponents, recent approaches using permutation entropy have shown a greater robustness of the complexity analysis against noise as inevitably present in experimental chaotic time-series[103]. Detailed mappings of the laser diode dynamics with optical feedback based on permutation entropy have identified those parameter regions where high complexity (large value of permutation entropy) and weak signature of laser parameters in the chaotic dynamics (in particular time-delay concealment) can be achieved simultaneously[102].

**Random number generation**

Although by nature distinct, chaos and randomness share a common feature in that they produce entropy. The Kolmogorov-Sinai entropy K is estimated from the sum of all positive Lyapunov exponents $\sum_i \lambda_i$ and quantifies the chaos sensitivity to initial conditions. The Shannon entropy $H(X) = \sum_i P(x_i) \log_2 P(x_i)$ measures the information content arising from uncertainty of the random variable $X$; $P(x_i)$ being the probability of the value $x_i$. For example, a fair coin toss has one bit of Shannon entropy since there are two possible outcomes (head and tail) that occur with equal probability. As commonly observed, any microscopic noise inevitably present in a chaotic system acts to amplify the divergence of nearby system trajectories in phase space. As time passes, the chaotic dynamics yields a Shannon information theoretic entropy directly related to the rate of growth of the divergence of system trajectories, i.e., to the Kolmogorov-Sinai entropy. Digitizing the chaotic output on 1 bit, the chaotic dynamics can ideally produce up to the maximum 1 bit of entropy as in fair coin tossing but at a rate up to the magnitude of the Lyapunov exponent.

Laser diode based chaos therefore makes an ideal physical source of random bits, since it combines unpredictability of the outcome with no dependence of the outcome on any previous outcome - two commonly accepted requirements for random number generation[104]. Also, chaotic laser diodes may produce a large number of positive Lyapunov exponents, whose magnitudes relate to laser frequencies that can be made very large. Random bits can therefore be produced at a much higher rate than other physical sources of entropy including quantum random number generators[105].

Since its first demonstration in 2008[106] the field of random number generation (RNG) using chaotic laser diodes has benefited from several developments. In the initial scheme [Fig. 4 a], use was made of a chaotic external-cavity laser diode. The feedback strength and the injection current are adjusted giving chaotic dynamics with a flat power spectrum of bandwidth about 10 GHz. A 1-bit analog-to-digital converter is used to produce the sequence of bits. However, sampling the chaotic output of a single laser diode did not produce an equal distribution of zeros and ones, i.e., this would not make a fair coin hence limiting the entropy. Use was then made of a second chaotic laser diode of the same type with their digitized outputs combined using an XOR logical operation. By making the time-delays of both external cavities and the sampling time of the ADC incommensurate, a 1.7 Gb/s sequence of bits that passed randomness statistical tests was achieved. Shortly afterwards, this scheme was simplified using only one chaotic laser diode and comparing its digitized output with a time-shifted version of it[107] (Fig 4b). Again, the time-shift, time-delay of the external-cavity and sampling clock time must be incommensurate to avoid any recurrence in the output.

Following that initial demonstration, several schemes were developed to improve the RNG performance, either focusing on the post-processing or on the physics of the laser diode chaos. The state-of-the-art is summarized in Figure 5.

Instead of using 1-bit ADC, benefit derives from the use of multi-bit extraction[108]. Keeping the $m$ least significant bits (LSBs) of the comparison between the digitized chaotic laser output and a time-shifted version (Fig. 4c), RNG at a rate of 12.5 Gb/s was demonstrated. The generation rate has naturally increased by considering more bits but also because the multi-bit extraction improves the symmetry of the distribution of zeros and ones, hence removing the inherent bias in the outcome. Instead of a single comparison (1st order derivative) the same group suggested the use of high-order derivatives[109]. Keeping $m$ LSBs of the $n^{th}$ order derivative (Fig. 4d) gives generation rates up to 300 Gb/s because the post-processing provides additional bits and improves the entropy growth rate, hence allowing for higher sampling rate.

Experiments have suggested that the performance of optical chaos-based RNG increases with improved flatness and bandwidth of the generated chaos[110]. Among the well-known techniques for bandwidth enhancement, using optical injection allows to translate spectrally the chaotic bandwidth of a master laser toward the relaxation oscillation frequency of the injected slave laser. Increasing the bandwidth of chaos up to about 16.5 GHz[107] has enabled the use of a 12,5 GS/s sampling rate on the 6 LSB output of the digitized chaos, i.e. 75 Gb/s RNG. The same setup but using reverse bit order sequence in the time-shifted laser output before applying the XOR logical operation uses the full 8 bit ADC resolution at the maximum sampling rate (50 GS/s) hence achieving the current record of 400 Gb/s[111].

Inspired by the previous work using chaotic laser diodes employing external optical feedback, numerous works have shown good RNG performance using integrated chaotic laser diodes. The same setup as the one used in 2008[106] has been utilized with DFB lasers integrated with a 1 cm long passive cavity[112]. The resulting flat power spectrum provides 2.08 Gb/s RNG using single bit extraction. A similar (1.56 Gb/s) performance has been achieved using a ring passive waveguide implementing optical feedback on a DFB laser with two SOA gain sections[60]. The use of multi-bit extraction with an integrated DFB laser with 1cm passive waveguide has been suggested to use 14 LSBs out of a 16 bit-ADC at a sampling rate of 10 GS/s hence improving the bit rate to 140 Gb/s [113].

Although optical feedback is an efficient technique for obtaining high bandwidth chaos with a large set of positive Lyapunov exponents, the time-delay periodicity imprinted in the laser output leads to recurrences in the outcome and the setup requires fine-tuning of the external cavity. A recent alternative was demonstrated using polarization chaos from a free-running VCSEL[114]: the 5 LSBs of the 8 bit-digitized polarized light output are used at a sampling rate of 20 GS/s, hence producing 100 Gb/s physical RNG successfully passing standard statistical tests for randomness. Promising results (up to 30Gb/s RNG) have also been obtained using the oversampling of 1.5 GHz low-pass filtered chaotic dynamics achieved by optical injection into a DFB laser diode[115].

**Chaotic optical sensing**

Chaotic lasers also find applications in high-precision ranging in so-called noise radar or correlation radar[116], or else chaotic radar[117] (CRADAR). The use of chaotic pulse trains in correlation ladar offers a high-bandwidth to enable high-precision range measurements. Rapid decorrelation due to the irregular pulse intervals and amplitudes yields unambiguous range measurements; improved signal-to-noise ratios are gained from the available high average pulse repetition frequencies. A proof of concept CRADAR system[117] using an optically injected semiconductor laser as the source of chaos achieved a range resolution of 9 cm limited by the detection bandwidth. Chaotic laser diodes have also been used to enhance resolution in optical time domain reflectometry (OTDR)[118]. OTDR is a key diagnostic tool e.g. for testing optical-fiber transmission systems. The challenges in OTDR are to increase the measurement range, enhance signal-to-noise ratios and improve spatial resolution. Using lasers driven into chaos via feedback from an optical fibre ring, a spatial resolution of 6 cm was achieved in distances in a 140m range[118]. Again, resolution was limited by the bandwidth of the detection.

**Optical Logic and Chaos Computing**

Chaos has also been proposed as a novel means for performing computing[119] and to implement optical logic functions[120]. A practical implementation of a NOR logic gate has been initially realized by applying a threshold function to the double-scroll chaotic attractor achieved in a Chua electronic circuit[120]. This proof-of-concept experiment demonstrates the universal computing capability of chaotic systems considering that all logic operations (AND, OR, NOT, XOR, NAND) can be constructed from combinations of NOR logic functions. Optoelectronic devices have later been used to implement such logic elements[121] including the use of chaotic two-section semiconductor lasers[122]. NOR logic operation is theoretically demonstrated by analyzing the synchronization properties of two mutually coupled chaotic two-section laser diodes with initially identical laser parameters. By modulating the injection currents in both the gain and absorber sections of one of the two laser diodes, similarly to what we defined previously as chaos shift keying (CSK), one achieves a situation where synchronization quality is high (output equal to 1) only when the gain and absorber bias currents of both lasers are almost identical (hence when both modulated current inputs are 0). The processing speed of the resulting NOR gate is however limited by the time for synchronization/desynchronization which typically takes values of several nanoseconds.

**Future outlook**

In order to assess how this area of activity may evolve in future it is appropriate to first reflect on the remarkable progress which has already been made. It is worth recalling some early scepticism amongst some members of the semiconductor laser community who considered that the complex dynamics of semiconductor lasers was just a hindrance to practical applications and so should be engineered out of existence. However, as appreciation grew of the universality of many nonlinear dynamical and chaotic phenomena it became apparent that the laser diode provided an ideal test-bed for investigations of novel dynamical behaviour. That attracted mathematicians and theoretical physicists who added to the insights and the armoury of techniques which could be used to explore laser diode dynamics. The richness of the dynamics which may be conveniently accessed in a variety of laser diode configurations augurs well that the field will remain a fertile area for exploration and exploitation for many years to come.

Exciting developments may be envisaged due to the continuing evolution of laser diode designs with particular opportunities arising with the demonstration of electrically pumped nano-lasers[123]. Such nano-laser designs may incorporate plasmonic and spintronic features which will impact the laser dynamics. Nano-lasers are being developed for applications in such diverse fields as quantum computing and systems-on-a-chip. Much attention has been given in recent years to developing semiconductor lasers operating in the mid-infra red and terahertz regions of the electromagnetic spectrum. The quantum cascade laser (QCL) whose operation relies on unipolar intersubband electronic transitions is the most prominent semiconductor laser operating in this wavelength range. Because of the technological challenges which have needed to be surmounted in order to create viable QCLs experimental investigation of their dynamical properties has been relatively limited but nevertheless exploration of nonlinear dynamical properties has begun[124]. As the operability of QCL develops one foresees opportunities for exploiting their novel dynamical features. Apart from working with existing semiconductor lasers, it is suggested that a fertile direction for development is the design of novel semiconductor lasers in order to exploit specific nonlinear dynamical phenomena. This cuts across the grain of much electronic engineering where operation in the linear regime is preferred but with the available wide range of nonlinear dynamical phenomena one can expect that useful applications can be satisfied by deliberately emphasising such aspects.

In terms of engineering applications of laser diode chaos a specific early focus was private optical communications. Particular mention may be made of efforts made within Europe to advance this technology. Although some of the momentum from that effort has been lost in recent years, there is evidence of continued interest in this topic elsewhere in the world and notably in China. Given the changes which are occurring in the global economy it may be the case that deployment of secure chaos-based communications systems will come to fruition driven by requirements of the growing economy in China.

Due to their ease of operation, laser diodes may be used to build experimental analogs of processes occurring in other fields. Thus for example time-delay effects in semiconductor lasers enable insights to be gained into synaptic behaviour in neurons[125]. As latency is a feature of many biological and physical systems it is expected that such analogies can be profitably used in many other cases. In some parameter ranges, laser diode dynamics with optical injection[126] or optical feedback[127] show similar extreme event statistics as those characterizing rogue waves in hydrodynamics[128]. The statistical analysis here takes advantage of the frequencies of the processes involved being much higher than in fluid dynamics. Another example is found in the study of dissipative systems. Many of the features of spatial dissipative solitons including chaotic motions and clustering of localized states[129] would benefit from the knowledge of coupled laser chaotic oscillators.

It is suggested that a particular legacy of the wide-ranging explorations of laser diode chaos is an enhanced awareness of powerful techniques for characterising and controlling complex dynamics. There is clear scope for wider application of such techniques. Currently global attention is being given to the development of low carbon economies. Photonics in general, and the laser diode in particular, has a direct role to play in that agenda with e.g. reducing energy consumption in optical communication networks[130] and efficient solid state lighting being key elements in the reduction of electricity consumption worldwide. However an indirect impact can also be identified. Increased use of localised sustainable sources of electricity will bring challenges in the control of complex distribution networks where nonlinear dynamical behaviour will arise. Insights derived from exploration of nonlinear and chaotic dynamics in laser diode systems may be directly applicable to tackling such engineering challenges.

[1]*Supélec, Laboratoire Matériaux Optiques, Photonique et Systèmes (LMOPS) (Supélec/Université de Lorraine) and Optics and Electronics (OPTEL) Research Group, 2 Rue Edouard Belin, 57070 Metz (France)*
[2]*Bangor University, School of Electronic Engineering, Dean Street, Bangor, LL57 1UT, Wales (UK)*

\* Corresponding author: marc.sciamanna@supelec.fr

### Acknowledgment

MS acknowledges the financial support provided by Fondation Supélec, Conseil Régional de Lorraine, Agence Nationale de la Recherche (ANR) through the project "TINO" grant number ANR-12-JS03-005, and the Inter-University Attraction Pole (IAP) program of BELSPO through the project "Photonics@be" grant number IAP P7/35. KAS acknowledges the financial support provided by the Sêr Cymru National Research Network in Advanced Engineering and Materials.

# Figure Captions

**Figure 1 - Chaos properties and chaos in lasers. a.** Reprint of the irregular pulsing dynamics observed in the output of the ruby solid-state laser (Ref. 1). **b.-e.** Numerical simulation of Lorenz chaos. The Lorenz equations (Ref. 3) are three nonlinear differential equations for *x, y, z* variables with parameters *r, b, σ*. We have fixed $r = 28$, $b = 8/3$, $\sigma = 10$. **b.** Time-traces of *x* for two slightly different initial conditions (black and red trajectories). **c.** Trajectories in the three-dimensional phase space with projections in the two-dimensional phase planes. **d.** Computation of the three Lyapunov exponents using the Wolf algorithm (Ref. 6) for the dynamics in b. The dynamics shows one positive Lyapunov exponent. **e.** Computation of the correlation dimension *D* using the Grassber-Procaccia algorithm (Ref. 7), which measures the dimensionality of the space occupied by a set of points. The algorithm computes the correlation integral $C(r)$ for increasing distance *r* between points and for increasing embedding dimension. For a large enough embedding dimension and in a given range of *r*, $C(r)$ scales as $C(r) = r^D$ where *D* is the correlation dimension that is a lower bound estimation of the fractal dimension. **f.** experimental observation of Lorenz chaos in a free-running NH3 laser [reprint from Hubner, U., Abraham, N.B., Weiss, C.O. *Phys. Rev. A* **40**, 6354-6365 (1989)]. Bottom panel: typical time-trace of the pulsating intensity. Right panel: computation of the correlation dimension from the experimental time-series. **g.** Simulated dynamics of a laser diode with an injection current step. The photon number and carrier inversion show damped oscillations toward a steady-state (relaxation oscillations). Dynamics is limited to spiralling relaxation flows in the phase plane (right).

**Figure 2 - Configurations for achieving chaos in laser diode. a.** Optical feedback. Top to bottom: from an external mirror, from a Fabry-Perot etalon defined by two mirrors M1 and M2 separated by a distance d, from a diffraction grating for mode selectivity, from a polarized optical feedback by placing e.g. a quarter-wave plate that rotates the polarization of the returning field. **b.** Optical injection from a master to a slave laser diode. **c.** External current modulation applied to a laser diode. **d.** Optoelectronic feedback by re-injecting a delayed and amplified signal from a photodiode that measures the laser output. **e.** Two-section laser diode with one section working as a gain or saturable absorber section depending on the applied voltage. **f.** Narrow stripe semiconductor laser with transverse loss modulation. **g.** Integrated chaotic laser diode with a passive feedback cavity and a gain/absorber section for adjusting the feedback strength. **h.** Integrated chaotic laser diode using mutual coupling between two DFB laser sections. **i.** Integrated chaotic laser diode with a ring passive cavity. **j.** Chaos resulting from mode competition, either polarization competition in VCSELs or competition between transverse modes in a broad area laser diode. **k.** Wavelength chaos generator using a nonlinear optoelectronic feedback on a DBR laser. A birefringent plate (BP) with two polarizers (P1 and P2) converts nonlinearly the incoming intensity. **l.** Intensity chaos generator using a nonlinear optoelectronic feedback with a Mach-Zender electro-optic modulator.

**Figure 3 - Chaos synchronization and chaos communication using laser diodes. a.** Schematic plot of a chaos synchronization experiment. An external-cavity laser diode (optical feedback from a mirror with a time-delay $\tau_1$) synchronizes its chaotic dynamics with a receiver laser. In case the receiver laser is also an external-cavity laser the configuration is termed "closed-loop". **b.** Synchronization diagram showing the receiver output power versus the emitter output power. Perfect synchronization would mean a straight line in this diagram. Both synchronization and anti-synchronization have been observed (reprint from Ref. 76). **c.** Mapping of the synchronization quality in the plane of the injection parameters (injection strength $R_{inj}$ versus frequency detuning $\Delta f$). A gray scale is used to quantify the synchronization error *σ*, with black meaning no error and thus a high synchronization quality (reprint from ref. 75). **d.** Schematic of the different possibilities for encoding a message $m(t)$ into a chaotic carrier generated by a laser diode. In chaos masking the message is simply added to the laser output. In chaos modulation the message adds to the laser output but also impacts the laser diode dynamics. In chaos shift keying the message is typically applied as a digital modulation of the driving current or any of the laser parameter, such that two distinct chaotic dynamics are generated for bits 0 and 1 of the message. Upon synchronization of the receiver, the message $m(t)$ is extracted by substracting the receiver output to the signal $s(t)$ injected to the receiver laser. **e.** Experimental realization of a message broadcasting using chaos synchronization of laser diodes [reprint from Lee, M.W. & Shore, K.A. *IEEE Photon. Tech. Lett.* **18**, 169-171 (2006)]. Panels (a) and (c) show the synchronization plot of receivers 1 and 2 with respect to transmitter, respectively. Panels (b) and (d) are the corresponding cross-correlation plots. **f.** First experimental realization of chaos communication on a fiber-optic network (reprint from Ref. 91). The message A is well decoded (C) at the receiver side although it appears completely masked in the transmitted signal (B). The BER however increases with increasing bit rate due to degradation of the synchronization quality.

**Figure 4 - Random number generation (RNG) using chaos from a laser diode. a.** Schematic of the first experimental realization using two external-cavity laser diodes and a XOR logical operation applied to their 1-bit digitized outputs. The clock sampling time $\tau_s$ and the two external-cavity delay times $\tau_1$, $\tau_2$ must be incommensurate to avoid recurrences in the random bits. The bottom panel shows the experimental result with the generation of a random bit sequence at 1.7 Gb/s bit rate (reprint from Ref. 106). **b-d.** Different realizations of the post-processing that improve the produced random bit rate.

**Figure 5 - State-of-the-art of the performances of RNG using chaos from a laser diode.** LD: laser diode, OF: optical feedback, OI: optical injection, SRL: semiconductor ring laser, VCSEL: vertical-cavity surface-emitting laser. LSB: least significant bit. The different realizations differ by either the system under investigation, the post-processing method, the number of bits, and/or the sampling rate.

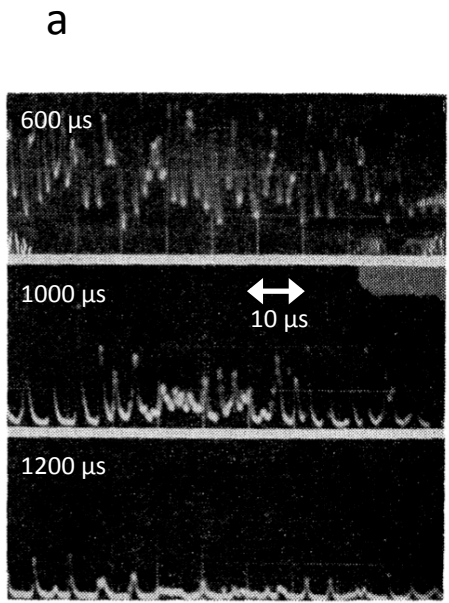
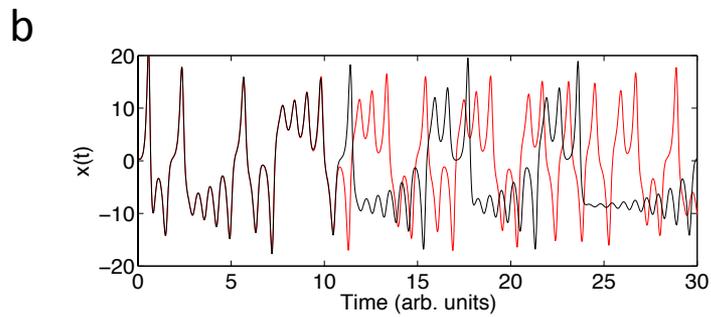
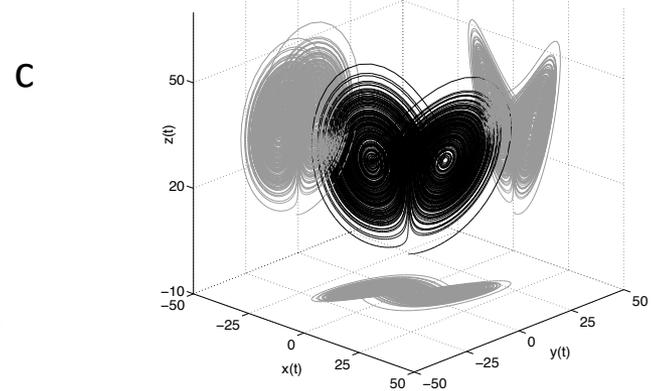
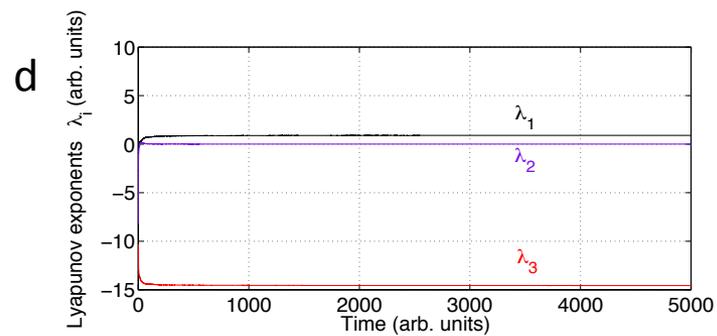
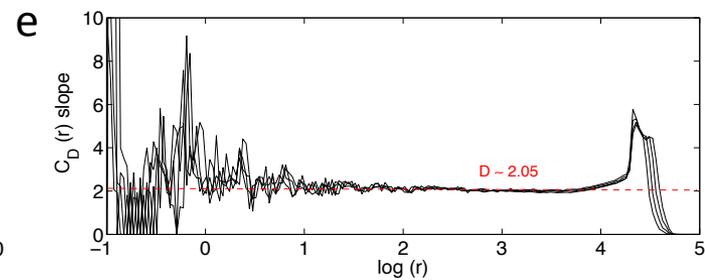
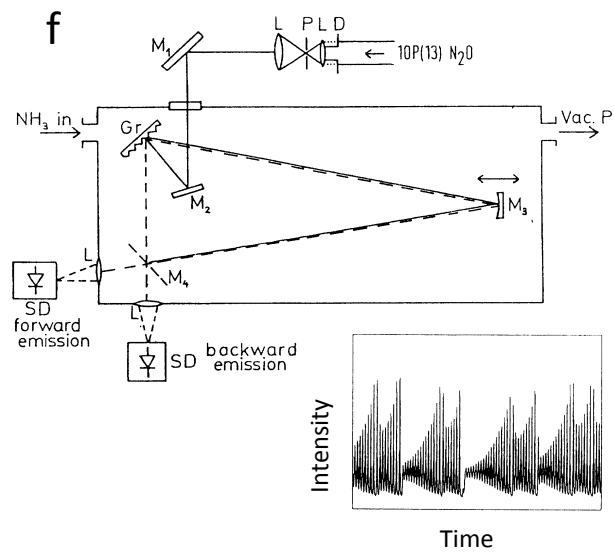
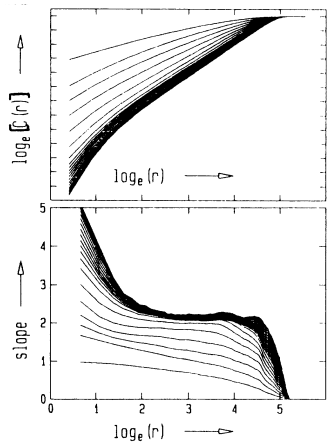
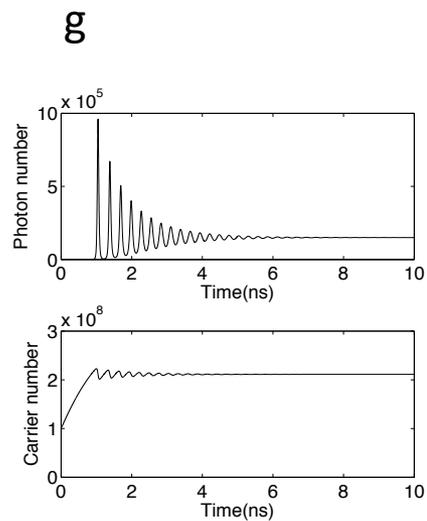
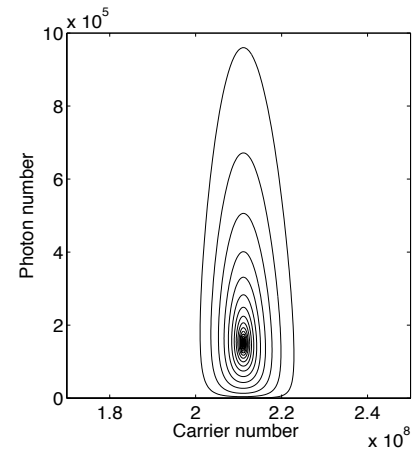

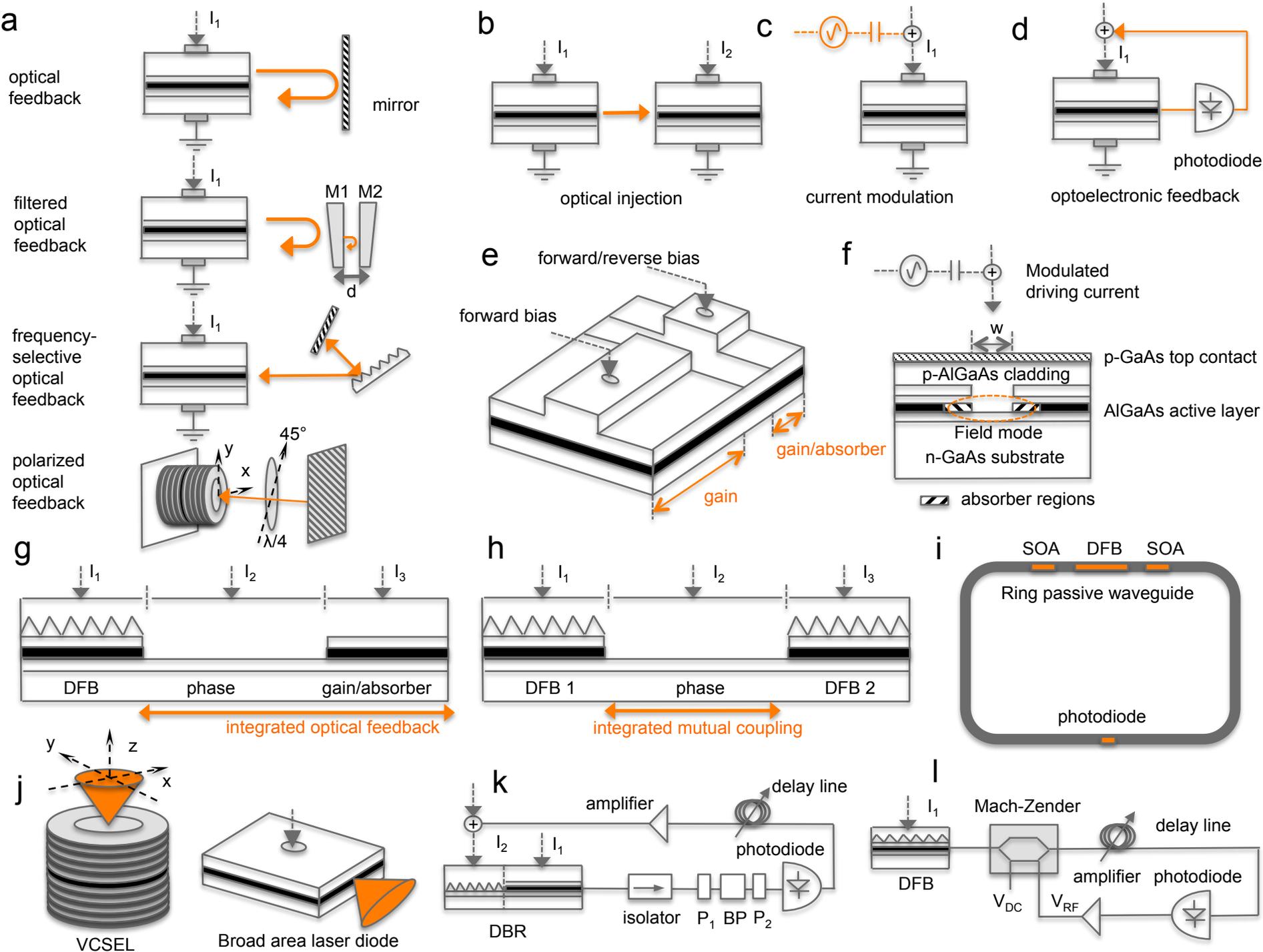

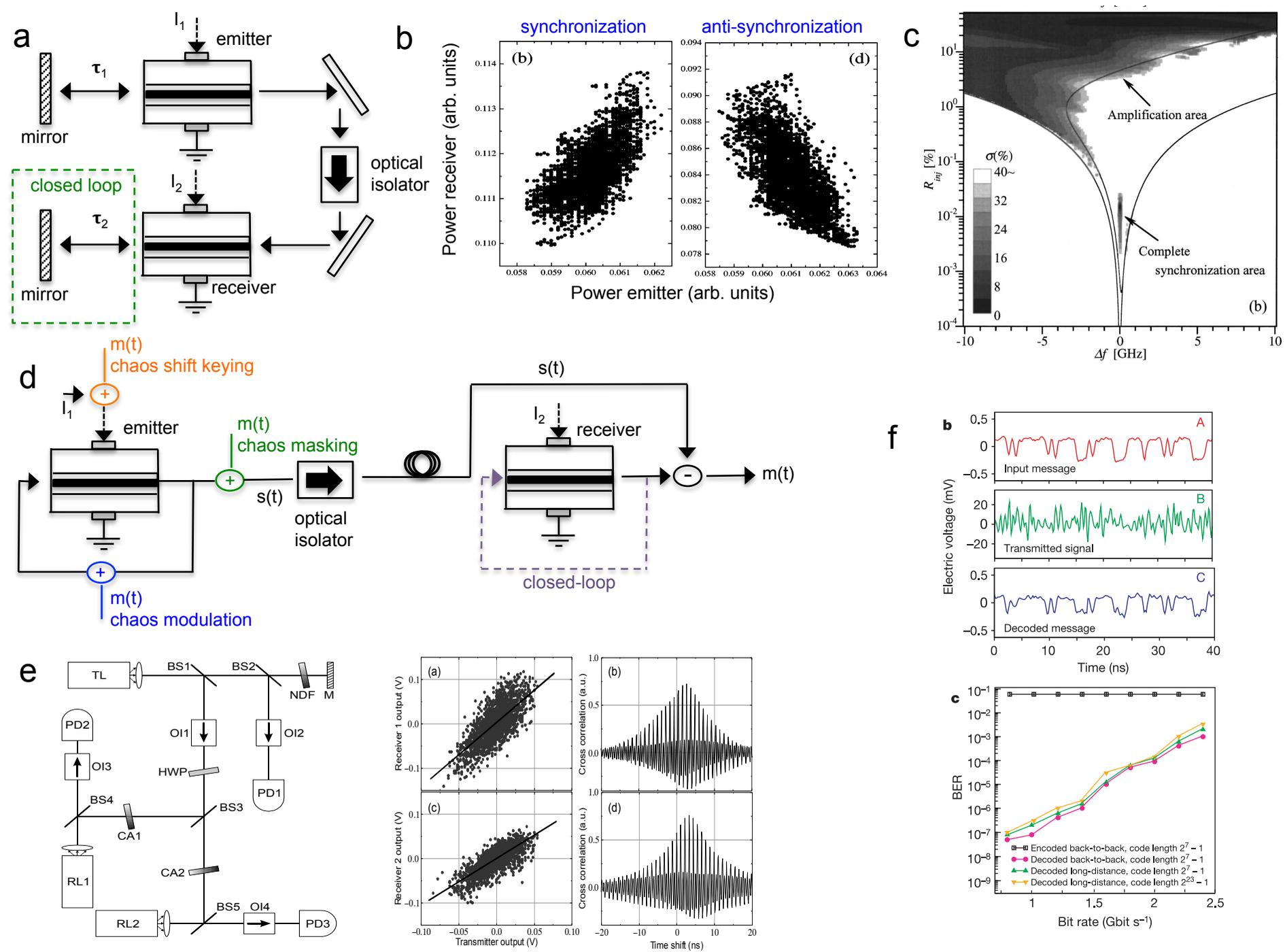

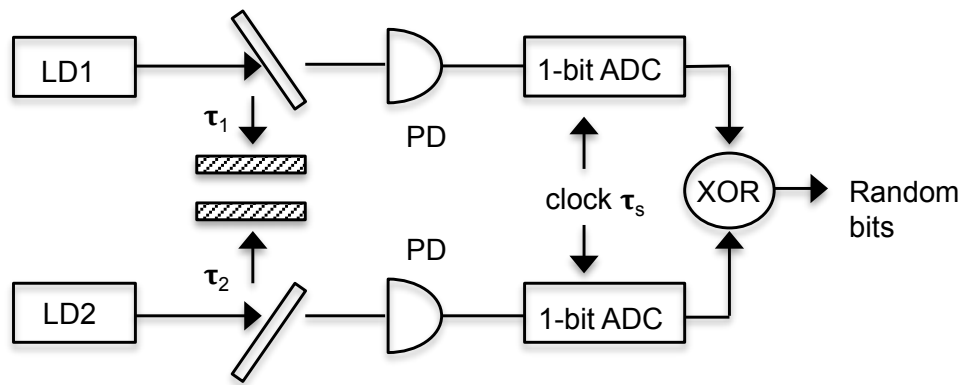
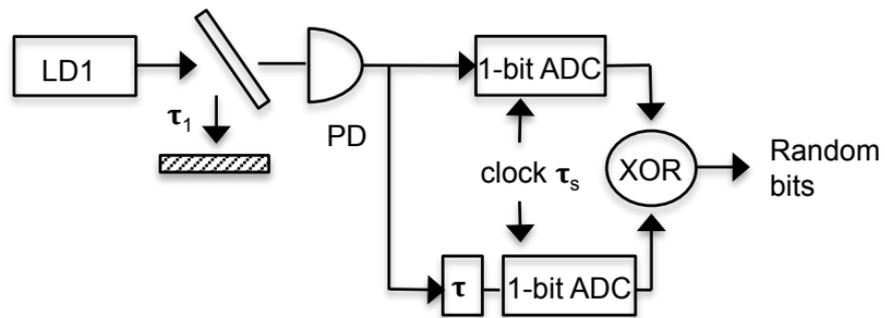
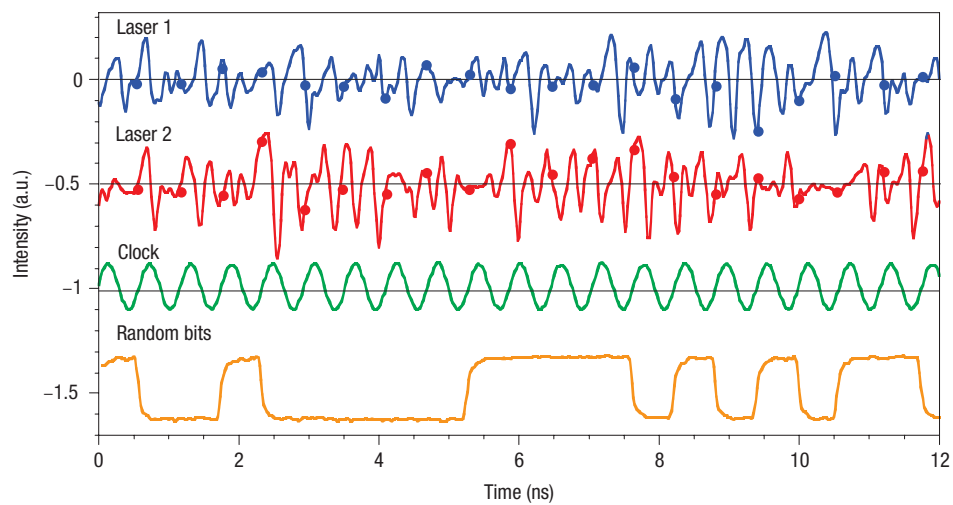
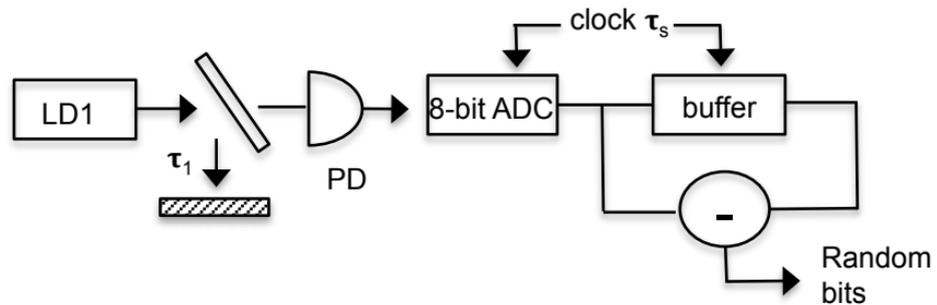
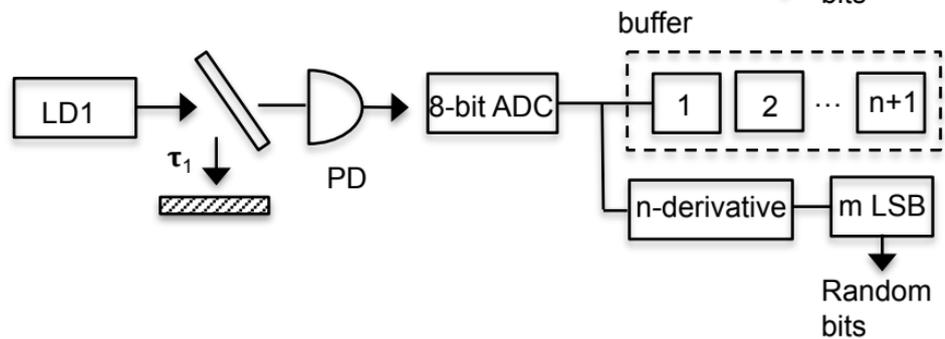

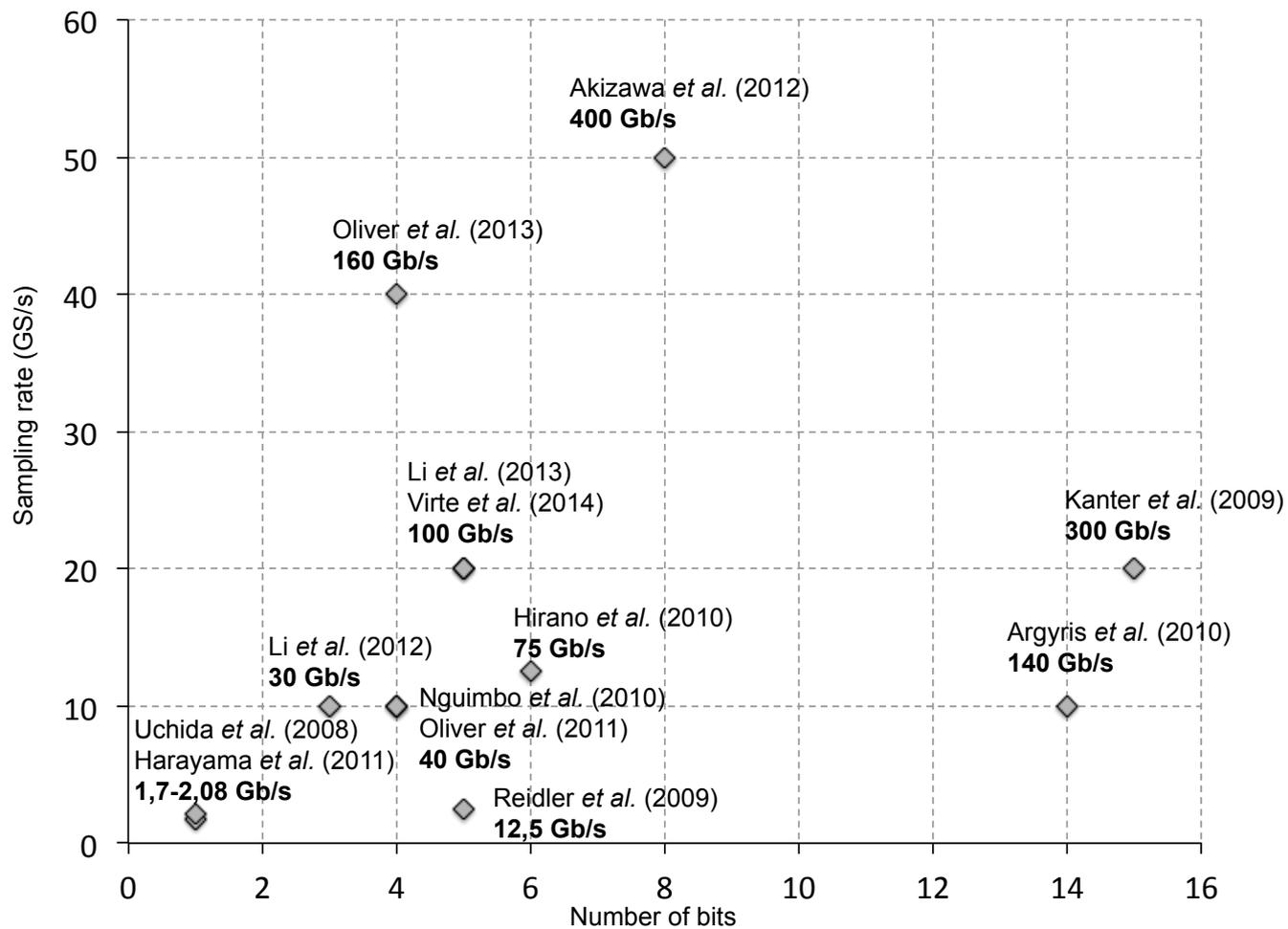

| Year | System | Post-processing | # bits | sampling rate (GS/s) | bit rate (Gb/s) | Reference |
|---|---|---|---|---|---|---|
| 2009 | (LD+OF) x2 | XOR | 1 | 1,7 | 1,7 | Uchida *et al.*, Nature Photon. 2, 728-732 (2008) |
| 2009 | LD+OF | 8-bit + derivative + LSB | 5 | 2,5 | 12,5 | Reidler *et al.*, Phys. Rev. Lett. 103, 024102 (2009) |
| 2009 | LD+OF | 8-bit + nth derivative + LSB | 15 | 20 | 300 | Kanter *et al.*, Nature Photon. 4, 58-61 (2010) |
| 2010 | (LD+OF) + (LD+OI) | 8-bit + XOR + LSB | 6 | 12,5 | 75 | Hirano *et al.*, Opt. Express 18, 5512-5524 (2010) |
| 2010 | LD+OF | 16-bit + LSB | 14 | 10 | 140 | Argyris *et al.*, Opt. Express 18, 18763-18768 (2010) |
| 2010 | SRL + OF | 8-bit x2 + XOR | 4 | 10 | 40 | Nguimdo *et al.*, Opt. Express 20, 28603-28613 (2012) |
| 2011 | (LD+OF) x2 | XOR | 1 | 2,08 | 2,08 | Harayama *et al.*, Phys. Rev. A 83, 031803(R) (2011) |
| 2011 | LD + OF (polarization) | 8-bit + LSB | 4 | 10 | 40 | Oliver *et al.*, Opt. Lett. 36, 4632-4634 (2011) |
| 2012 | (LD+OF) + (LD+OI) | 8-bit + bit reverse + XOR | 8 | 50 | 400 | Akizawa *et al.*, IEEE Photonics Technol. Lett. 24, 1042-1044 (2012) |
| 2012 | LD+OI | 8-bit + time-shift + XOR + LSB | 3 | 10 | 30 | Li *et al.*, Opt. Lett. 37, 2163-2165 (2012) |
| 2013 | LD + OF (polarization) | 8-bit + LSB | 4 | 40 | 160 | Oliver *et al.*, IEEE J. Quantum Electron. 49, 910-918 (2013) |
| 2013 | LD + OI | 8-bit + time-shift + XOR + LSB | 5 | 20 | 100 | Li *et al.*, IEEE J. Quantum Electron. 49, 829-838 (2013) |
| 2014 | VCSEL | 8-bit + time-shift + LSB | 5 | 20 | 100 | Virte *et al.*, Opt. Express 22, 17271-17280 (2014) |